\documentclass[conference]{IEEEtran}
\IEEEoverridecommandlockouts

\makeatletter
\def\markboth#1#2{\def\leftmark{\@IEEEcompsoconly{\sffamily}\MakeUppercase{\protect#1}}%
\def\rightmark{\@IEEEcompsoconly{\sffamily}\MakeUppercase{\protect#2}}}
\makeatother

\usepackage{graphicx}
\usepackage{epsfig}
\usepackage{epstopdf}
\usepackage{amstext}
\usepackage{amssymb}
\usepackage{amsmath}
\usepackage{amsthm}
\usepackage[font=footnotesize]{caption}
\usepackage{subcaption}
\usepackage{cite}
\usepackage{float}
\usepackage{array}
\usepackage{url}
\usepackage{flexisym}
\usepackage{soul}
\usepackage{color}
\usepackage{bbm}
\usepackage{dsfont}
\usepackage[T1]{fontenc}
\usepackage[utf8]{inputenc}
\usepackage{authblk}
\usepackage{acronym}
\usepackage[ruled,vlined,linesnumbered]{algorithm2e}
\usepackage{lipsum}  

\def\antboresightgain{G_{\rm B}}
\def\antsidegain{G_{\rm S}}
\def\boresightangle{\theta_{\rm B}}
\def\sideangle{\theta_{\rm S}}

\def\orientationTX{\phi_{\rm tx}}
\def\orientationRX{\phi_{\rm rx}}
\def\attenuationTXRX{A}
\def\residualSelfInterference{I_{\rm SI}}

\IEEEoverridecommandlockouts

\begin{document}
\newacro{PPP}[SPPP]{spatial Poisson point process}
\newacro{BS}[BS]{base station}
\newacro{FD}[FD]{full duplex}
\newacro{HD}[HD]{half duplex}
\newacro{CCDF}[CCDF]{complementary cumulative distribution function} 
\newacro{SNR}[SNR]{signal-to-noise ratio}
\newacro{SINR}[SINR]{signal-to-interference-plus-noise ratio}
\newacro{PDF}[PDF]{probability density function} 
\newacro{ASE}[ASE]{area spectral efficiency} 

\title{Area Spectral Efficiency and Coverage for Mixed Duplexing Networks with Directional Transmissions\vspace{-2mm}}

\author{Sanjay Goyal}
\author{Alphan \c{S}ahin}
\author{Robert L. Olesen \vspace{-2mm}}
\affil{InterDigital Communications, Inc., Melville, New York, USA \vspace{-3mm} }
\affil{\{Sanjay.Goyal, Alphan.Sahin, Robert.Olesen\}@InterDigital.com \vspace{-2mm}}
\maketitle
%
\begin{abstract}
In this paper, we consider a system of small cells assuming full duplex (FD) capable base stations (BSs) and half duplex (HD) user equipment (UEs). We investigate a mixed duplexing cellular system composed of FD and HD cells, when BSs are using directional transmissions. A stochastic geometry based model of the proposed system is used to derive the coverage and area spectral efficiency (ASE) of both BSs and UEs. The effect of FD cells on the performance of the mixed system is presented under different degree of directionality at the BSs. We show that enabling directional transmissions at the BSs yields significant ASE and coverage gain in both downlink and uplink directions. With directional transmissions, the ASE increases rapidly with the number of FD cells while the drop in the coverage rate due to  FD operations reduces significantly.

\end{abstract}


\section{Introduction}\label{sec1}

Beam-centric design is expected to be one of the key concepts for achieving higher throughput and spectral efficiency in the Fifth Generation (5G) of cellular networks \cite{NGMN_5G,beam_5G}. In this concept, directional transmissions are employed to increase the received signal quality at the intended users. As the signal energy is concentrated in a narrow region with directional transmissions, this helps to increase \ac{SNR} for a given link distance while reducing the interference among the users. 

Another key technology considered for 5G wireless systems is \ac{FD} communication with simultaneous transmission and reception on the same carrier. 5G wireless systems are being developed by the 3GPP New Radio (NR) standardization activity to meet performance requirements for IMT-2020. 3GPP has decided that NR will support paired and unpaired spectrum using frequency (FDD) and time (TDD) division duplexing operations, and will strive to maximize commonality between the technical solutions~\cite{3GPP:7}, allowing a flexible duplexing. The ability to assign transmission resources simultaneously to different transmission directions will allow efficient utilization of the available spectrum, enable future FD solutions.
Albeit its potential benefits, it has been reported that simultaneous downlink and uplink transmissions increase the interference floor in a network, introduce a trade-off between \ac{ASE} and coverage  \cite{Goyal_arxiv,Goyal_CommMag, SGoyal_ICC_Arxiv}. In this study, our goal is to provide further insights on this trade-off due to \ac{FD} communications  when directional transmissions are employed at the \acp{BS} by considering the beam-centric design philosophy of the 5G networks.

In literature, FD operation in wireless networks have been investigated considering different scenarios \cite{Haenggi_FD, Quekhybrid, alves2014average, elsawy_alpha_duplex}. While Tong \emph{et al.}~\cite{Haenggi_FD} analyze the throughput of a wireless network with FD radios using stochastic geometry in an ad-hoc setting, Lee \emph{et al.}~\cite{Quekhybrid} derive the throughput of a mixed  network considering only downlink and/or FD \acp{BS}. Alves \emph{et al.}~\cite{alves2014average} show the impact of residual self-interference on the spectral efficiency for a dense network along with FD operation. In \cite{elsawy_alpha_duplex}, the authors propose a scheme which allows a partial overlapping between uplink and downlink bands to maximize the gains with FD operations in each cell. Nevertheless the papers \cite{alves2014average, Quekhybrid, elsawy_alpha_duplex} mentioned above assume that the user equipment (UEs) to have FD capabilities, which is not practical given existing FD circuit designs \cite{survey_JSAC}. In addition, these studies investigate the ASE without assessing the outage probability in the network. In our earlier work~\cite{SGoyal_ICC_Arxiv}, we analyze the performance of mixed duplexing cellular systems, i.e., mixed system, composed of FD and \ac{HD} \acp{BS} with omni-directional antennas. We show that the fraction of FD cells can be used as a design parameter to target different ASE vs. coverage trade-offs; in particular, by increasing the amount of FD cells in the mixed system, the overall ASE increases at the cost of a drop in terms of coverage, and vice-versa. Psomas \emph{et al.}~\cite{directional_FD} quantify the impact of directionality in FD cellular networks, where all the \acp{BS} are in FD mode. The case of UEs with FD capability is compared against the case of UEs with only \ac{HD} capability, where the latter case is shown to have more potential from both performance as well as practical implementation perspectives. Among the existing papers addressing FD for wireless networks in cellular systems, to the best of our knowledge, there is no comprehensive study that addresses the ASE vs. coverage trade-off in mixed systems for the uplink and the downlink, with directional transmissions at the \acp{BS}, available. 


In this paper, we consider a mixed system where the \acp{BS} are using directional transmissions. A stochastic geometry-based model is utilized to investigate the impact of directional transmissions on the performance of a mixed system. We derive \ac{SINR} for both uplink and downlink by taking the impact of all intra and inter-cell interference. We also provide a model to calculate residual self-interference under FD operation with directional transmission. In particular, we analyze the \ac{ASE} vs. coverage trade-off of the mixed system as a function of the proportion of FD cells under different degree of directionality at the \acp{BS}. Among our main findings, we show that the trade-off between \ac{ASE} and coverage due to FD communications decreases when directional transmissions are employed at the \acp{BS}, i.e., increasing the number of \ac{FD} cells increases the \ac{ASE} significantly with a small loss in the coverage of the network.

The remainder of this paper is organized as follows. In Section~\ref{sec:Model}, we describe the system model. We show our formulation for computing the SINR and ASE for both downlink and uplink directions in Section~\ref{sec:MainSINR} and Section~\ref{sec:avg_rate}, respectively. In Section~\ref{sec:num_res}, we present and discuss the results while the conclusions are drawn in Section~\ref{sec:conc}.

\section{System Model}\label{sec:Model}
In this section, we describe the mathematical models for \ac{BS} deployment,  directional transmission, residual self-interference for \ac{FD} communications with directional transmissions, wireless channel, and power control.

\subsection {Deployment and Duplexing}\label{sec:deploy}
We consider a network where \acp{BS} are distributed according to a homogeneous and isotropic \ac{PPP} $\Phi_{\mathrm{B}}$ with density $\lambda_{\mathrm{B}}$. We assume that the \acp{BS} are capable of both \ac{HD} and \ac{FD} modes, while the UEs are limited to only \ac{HD} operation. The probabilities of a \ac{BS} to be in \ac{FD} mode, downlink \ac{HD} mode, and uplink \ac{HD} mode are denoted by $\rho_{\mathrm{F}}$, $\rho_{\mathrm{D}}$, and $\rho_{\mathrm{U}}$, respectively, where $\rho_{\mathrm{F}}+\rho_{\mathrm{D}}+\rho_{\mathrm{U}}=1$. Based on {\em Thinning theorem} \cite{haenggi2013stochastic}, the locations of the \acp{BS} in \ac{FD}, downlink \ac{HD}, and uplink \ac{HD} modes  can be modeled as independent \acp{PPP} denoted by $\Phi_{\mathrm{B}}^{\mathrm{F}}$, $\Phi_{\mathrm{B}}^{\mathrm{D}}$, and $\Phi_{\mathrm{B}}^{\mathrm{U}}$, where the corresponding densities of the \acp{PPP} are $\rho_{\mathrm{F}} \lambda_{\mathrm{B}} $, $\rho_{\mathrm{D}} \lambda_{\mathrm{B}}$, and  $\rho_{\mathrm{U}} \lambda_{\mathrm{B}}$, respectively. 

Each UE is assumed to be served by the nearest \ac{BS}, which leads to a Voronoi tessellation where the generators of the tessellation are the \ac{BS} locations, and the distribution of UE location is uniform in a Voronoi cell. We further assume that each \ac{BS} in \ac{FD} mode serves one uplink UE and one downlink UE on the same resources while each \ac{BS} in \ac{HD} mode communicates with one UE in the either downlink or uplink direction, i.e., UEs are always in \ac{HD} mode. We denote the set of downlink and uplink UEs served by the FD \acp{BS} on the same resource  as ${\Phi}_{\mathrm{U}}^{\mathrm{F,D}}$ and ${\Phi}_{\mathrm{U}}^{\mathrm{F,U}}$, respectively. Similarly,  the set of downlink and uplink UEs served by the \ac{HD} \acp{BS} on the same resource are expressed as ${\Phi}_{\mathrm{U}}^{\mathrm{H,D}}$ and ${\Phi}_{\mathrm{U}}^{\mathrm{H,U}}$, respectively. It is worth noting that ${\Phi}_{\mathrm{U}}^{\mathrm{F,D}}$, ${\Phi}_{\mathrm{U}}^{\mathrm{F,U}}$, ${\Phi}_{\mathrm{U}}^{\mathrm{H,D}}$, and ${\Phi}_{\mathrm{U}}^{\mathrm{H,U}}$ are not \acp{PPP} since the UEs are associated with the nearest \ac{BS}. Nevertheless, we model these subsets as \ac{PPP} for the sake of tractable analysis, which has also been considered in \cite{jeffAndrews_uplink,dynamic_TDD,SGoyal_ICC_Arxiv} as this approximation yields well-aligned \ac{SINR} distributions. We can obtain the densities of ${\Phi}_{\mathrm{U}}^{\mathrm{F,D}}$, ${\Phi}_{\mathrm{U}}^{\mathrm{F,U}}$, ${\Phi}_{\mathrm{U}}^{\mathrm{H,D}}$, and ${\Phi}_{\mathrm{U}}^{\mathrm{H,U}}$ as  $\lambda_{\mathrm{U,F,D}} = \rho_{\mathrm{F}}\lambda_{\mathrm{B}}$, $\lambda_{\mathrm{U,F,U}} = \rho_{\mathrm{F}}\lambda_{\mathrm{B}}$, $\lambda_{\mathrm{U,H,D}} = \rho_{\mathrm{D}}\lambda_{\mathrm{B}}$, and
$\lambda_{\mathrm{U,H,U}} = \rho_{\mathrm{U}}\lambda_{\mathrm{B}}$, respectively, by exploiting {Thinning theorem} \cite{haenggi2013stochastic}. In addition, we assume that the set ${\Phi}_{\mathrm{U}}$ of all  UEs, which is the union ${\Phi}_{\mathrm{U}}^{\mathrm{F,D}} \cup {\Phi}_{\mathrm{U}}^{\mathrm{F,U}} \cup {\Phi}_{\mathrm{U}}^{\mathrm{H,D}} \cup {\Phi}_{\mathrm{U}}^{\mathrm{H,U}}$; ${\Phi}_{\mathrm{U}}$ is an \ac{PPP} and its density is the sum of each subset's density, which is~($\rho_{\mathrm{F}} +1) \lambda_{\mathrm{B}}$~\cite{baccellistochastic}. ${\Phi}_{\mathrm{U}}^{\mathrm{F,D}}$, ${\Phi}_{\mathrm{U}}^{\mathrm{F,U}}$, ${\Phi}_{\mathrm{U}}^{\mathrm{H,D}}$, and ${\Phi}_{\mathrm{U}}^{\mathrm{H,U}}$ are also assumed to be independent of one another and independent of $\Phi_{\mathrm{B}}^{\mathrm{F}}$, $\Phi_{\mathrm{B}}^{\mathrm{D}}$, and $\Phi_{\mathrm{B}}^{\mathrm{U}}$ to maintain  model tractability, which are also considered in the previous work \cite{SGoyal_ICC_Arxiv,alves2014average, elsawy_alpha_duplex}.

\begin{figure*}[t!]
	
    \centering
    \begin{subfigure}[b]{0.19\textwidth}
        \centering
		\includegraphics[width = 1.3 in] {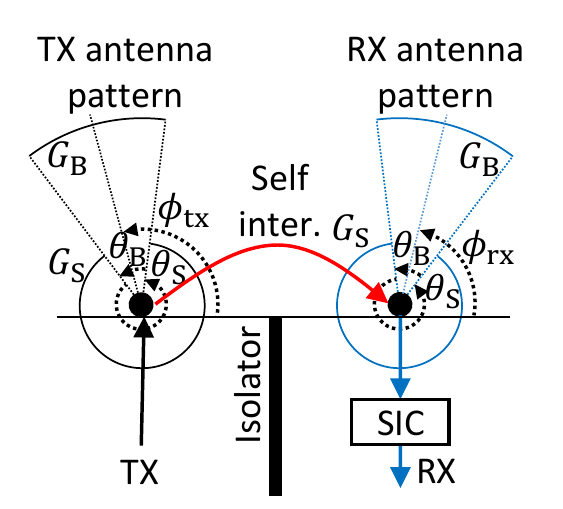}
        \caption{FD model with directional antennas at BS.}
         \label{fig:SIC}
    \end{subfigure}~  
    \begin{subfigure}[b]{0.19\textwidth}
        \centering
		\includegraphics[width = 1.1 in] {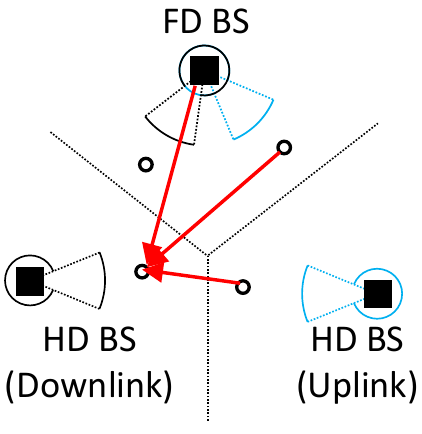}
        \caption{Interference at HD cell (downlink).}
         \label{fig:dlc2}
    \end{subfigure}~
    \begin{subfigure}[b]{0.19\textwidth}
        \centering
		\includegraphics[width = 1.1 in] {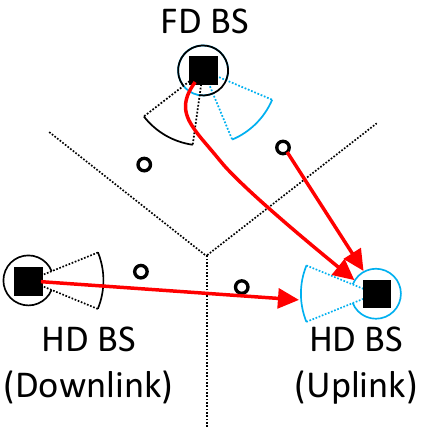}
        \caption{Interference at HD cell (uplink).}
         \label{fig:ulc2}
    \end{subfigure}~ 
    \begin{subfigure}[b]{0.19\textwidth}
        \centering
		\includegraphics[width = 1.1 in] {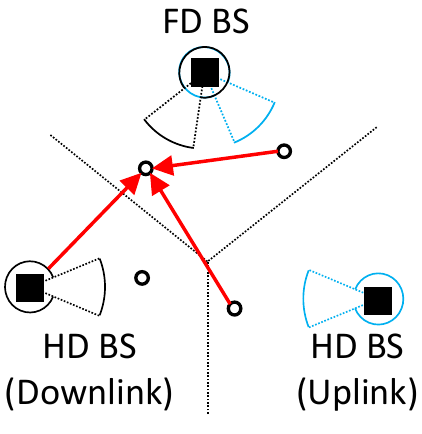}
        \caption{Interference at FD cell (downlink).}
         \label{fig:dlc1}
    \end{subfigure}~    
    \begin{subfigure}[b]{0.19\textwidth}
        \centering
		\includegraphics[width = 1.1 in] {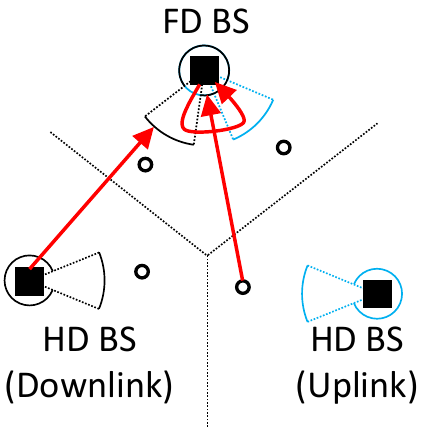}
        \caption{Interference at FD cell (uplink).}
        \label{fig:ulc1}
    \end{subfigure}~
    \caption{Interference cases in mixed FD and HD scenarios in downlink and uplink, and FD model with directional transmission at BSs.}
\label{fig:systemModel}
\vspace{-3mm}
\end{figure*}

\subsection {Directional Transmission}\label{sec:dir}
We consider directional antennas at the \acp{BS} and omnidirectional antennas at the UEs. For \ac{FD} communications, we assume that each \ac{BS} employs two directional antennas; one for the uplink and one for the downlink, and each of them is steerable to an independent direction. For the sake of tractability, we consider a directional antenna model which characterizes the main lobe and the side lobes of the antenna pattern with single variables as
\begin{equation}\label{eq:antennaPattern}
	G(\theta) =
	\left \{ \,
		\begin{IEEEeqnarraybox}[] [c] {l?s}
			\IEEEstrut
			 \antboresightgain  &  if $\theta\le|\boresightangle/2|$, \\
			\antsidegain &  if $\theta\ge|\boresightangle/2|$,
			\IEEEstrut
		\end{IEEEeqnarraybox}
	\right.
\end{equation} 
where $\theta$ is the radiation angle in local coordinate systems of the antenna, $\antboresightgain$ is the antenna gain for the main lobe of the antenna radiation pattern which spans $\boresightangle$ degrees and $\antsidegain$ is the antenna gain for the side lobe of the antenna radiation pattern which spans the rest of the angles, i.e., $\sideangle=2\pi-\boresightangle$ degrees as illustrated in \figurename\ref{fig:systemModel}\subref{fig:SIC}. It is worth noting that various antenna models are proposed for system-level investigations in the literature (e.g., \cite{3GPP:5, 3GPP:6, Gunnarsson2008}  and the references therein).  It is possible to integrate these mathematical models in our derivations in the following sections or approximate them by introducing multiple gain levels to \eqref{eq:antennaPattern}. Since our goal in this study is to focus on the insights for the coverage and \ac{ASE} in a mixed FD and \ac{HD} network with a directional antenna setting at the \acp{BS} via a tractable approach, we provide our results by considering the model given in \eqref{eq:antennaPattern}.

\subsection {Residual Self-Interference in Full Duplex Mode}
\label{sec:selfInterference}
We consider a self-interference between transmit and receive branches in FD mode which takes directional antenna model described in Section \ref{sec:dir}, as illustrated in \figurename\ref{fig:systemModel}\subref{fig:SIC}. Since we assume that the directional antennas at the receive and transmit chains are steerable independently, i.e., one direction for uplink UE and another direction for downlink UE in FD \ac{BS}, the amount of the self-interference is a function of the antenna orientations. For example, if uplink UE and downlink UE are on the opposite sides relative to the \ac{BS} location, the amount of the interference can be  amplified or attenuated significantly due to the the antenna gains depending on the orientations of the transmit and receive antennas. We model the residual self-interference 
$\residualSelfInterference$ as
\begin{equation}\label{eq:selfInterference}
\small 
	\residualSelfInterference =
	\left \{ \,
		\begin{IEEEeqnarraybox}[] [c] {l?s}
			\IEEEstrut
			\attenuationTXRX P_{\rm B}\antboresightgain^2  &  if $C_1^{\rm SI}:|\orientationTX|\le\boresightangle/2$, $ |\pi-\orientationRX|\le\boresightangle/2$, \\
            \attenuationTXRX P_{\rm B}\antboresightgain\antsidegain  & if  $C_2^{\rm SI}:|\orientationTX|\ge\boresightangle/2$, $ |\pi-\orientationRX|\le\boresightangle/2$, or\\
              &  $|\orientationTX|\le\boresightangle/2$, $ |\pi-\orientationRX|\ge\boresightangle/2$, \\
            \attenuationTXRX P_{\rm B}\antsidegain^2  & if $C_3^{\rm SI}:|\orientationTX|\ge\boresightangle/2$, $ |\pi-\orientationRX|\ge\boresightangle/2$,
			\IEEEstrut
		\end{IEEEeqnarraybox}  
	\right.
\end{equation}
where  $\orientationTX$ and $\orientationRX$ are the transmit and receive antenna boresights in local coordinates of the \ac{BS}, respectively, and $\attenuationTXRX$ is the effective self-interference cancellation gain which is a function of path loss, analog, and digital cancellation mechanisms at the \ac{BS}.

\subsection{Wireless Channel and Power Allocation}\label{sec:ChModel}
We consider different wireless channel models between the devices considering the device type, which is also recommended by 3GPP for \ac{BS}-to-\ac{BS}, \ac{BS}-to-UE, and UE-to-UE links \cite{3GPP:1}. Without loss of generality, we assume that the path loss model between \ac{BS} and UE, UEs, and \acp{BS} are $\mathrm{PL}_{1}(d)=K_{1}d^{-\alpha_{1}}$, $\mathrm{PL}_{2}(d)=K_{2}d^{-\alpha_{2}}$, and $\mathrm{PL}_{3}(d)=K_{3}d^{-\alpha_{3}}$, respectively, where $\alpha_{1}$, $\alpha_{2}$, and $\alpha_{3}$ are the path loss exponents, and $K_{1}$, $K_{2}$, and $K_{3}$ are the signal attenuations at distance $d$ = 1. We consider that the transmitted signal is exposed to Rayleigh fading, which leads to exponentially distributed received signal power $\sim$ exp($\mu$) with mean $\mu^{-1}$. 

In the downlink, the \ac{BS} transmission powers are assumed to be identical and denoted as $P_{\mathrm{B}}$. For the uplink, we consider fractional power control where each UE, which is at distance $R$ from its serving \ac{BS} transmits with power  $ P_{\mathrm{U}} K_{1}^{-\epsilon }R^{\epsilon \alpha_{1}}$, where $\epsilon \in [0,1]$ is the power control factor \cite{jeffAndrews_uplink}.

\section{SINR Distributions}\label{sec:MainSINR}
In this section, we provide the analytic expressions of \ac{SINR} \acp{CCDF} and \ac{ASE} in downlink and uplink. 
Since our system model considers a mixed network which includes \acp{BS} in \ac{FD} mode and \acp{BS} in \ac{HD} in both uplink and downlink directions at the same time, we first evaluate \ac{SINR} \acp{CCDF} in \ac{FD} cell and \ac{HD} cell separately. Subsequently, we derive \ac{ASE} and coverage rate considering the complete network.

Regardless of the duplexing mode, all of the active downlink \acp{BS} and all of the active uplink UEs in other cells interfere the intended link in both downlink and uplink.
In a \ac{HD} cell, both uplink UEs in the neighboring \ac{FD} cells and their corresponding \acp{BS} interfere the intended received signal; that is either in downlink or in uplink as illustrated in \figurename\ref{fig:systemModel}\subref{fig:dlc2} and  \figurename\ref{fig:systemModel}\subref{fig:ulc2}, respectively. On the other hand, in the downlink of an FD cell, the uplink UE interferes the downlink signal for the UE located in the same FD cell as shown in \figurename\ref{fig:systemModel}\subref{fig:dlc1}. In the uplink of an FD cell, the \ac{BS} interferes itself based on the model given in \eqref{eq:selfInterference}, as shown in \figurename\ref{fig:systemModel}\subref{fig:ulc1}.

\subsection{Downlink SINR Distribution in an FD Cell}\label{sec:FDdownlinkSINR}
The downlink SINR at a UE of interest in an FD cell can be expressed as 
\vspace{-0mm}
\begin{equation}\label{eq:sinr_fd_dwn}
\gamma_{\mathrm{FD,UE}}=\frac{P_{\mathrm{RX,UE}}}{N_{0}+I_{\mathrm{D}}+I_{\mathrm{U}}},
\vspace{-0mm}
\end{equation}
where $N_{0}$ is the noise power at the UE, and $P_{\mathrm{RX,UE}}$ is the received signal power from the serving \ac{BS}, given by
\vspace{-0mm}
\begin{equation}\label{eq:p_rx}
P_{\mathrm{RX,UE}} = P_{\mathrm{B}} \antboresightgain g_{b_{0}} K_{1}r^{-\alpha_{1}},
\vspace{-0mm}
\end{equation}
where $r$ is the distance between the UE and its serving \ac{BS}. 
The serving \ac{BS} is indicated by $b_0$, and $g_{b_{0}}$ denotes the Rayleigh fading affecting the signal from the \ac{BS} $b_0$. $I_{\mathrm{D}}$ and $I_{\mathrm{U}}$ are the total interference received at the UE from all the downlink transmissions and from all the uplink transmissions, respectively. 

The total interference from all the downlink transmissions including all FD cells ($\Phi_{\mathrm{B}}^{\mathrm{F}} \backslash{b_0}$) and all \ac{HD} downlink cells ($\Phi_{\mathrm{B}}^{\mathrm{D}}$) can be defined as
\begin{equation}\label{eq:int_D_A}
\begin{aligned}
I_{\mathrm{D}} = \sum_{b      \in \{\Phi_{\mathrm{B}}^{\mathrm{D}}       \cup       \Phi_{\mathrm{B}}^{\mathrm{F}} \backslash{b_0}\}} G_b^{\rm{DD}} P_{\mathrm{B}}{g_b K_{1}R_b^{-\alpha_{1}}},
\end{aligned}
\end{equation}
where $R_b$ is the distance between the UE of interest and the interfering  \ac{BS} $b$, and $g_{b}$ denotes the Rayleigh fading for this link, and $G_b^{\rm{DD}}$ is the effective antenna gain expressed as
\begin{equation}\label{eq:int_D_def}
\small{
	G_b^{\rm{DD}} = 
	\left \{ \,
		\begin{IEEEeqnarraybox}[] [c] {l?s}
			\IEEEstrut
			 \antboresightgain  &  if $\rm{C}_1^{\rm{DD}}$ holds, \\
			 \antsidegain  &  if $\rm{C}_2^{\rm{DD}}$ holds,
			\IEEEstrut
		\end{IEEEeqnarraybox}
	\right. 
	}
\end{equation}
where $\rm{C}_1^{\rm{DD}}$ and  $\rm{C}_2^{\rm{DD}}$  are the conditions that the transmit beam of the interfering \ac{BS} $b$ is or is not oriented towards the UE of interest, which occurs with the probability of $p_1^{\rm{DD}} = \boresightangle/2\pi$ and the probability of $p_2^{\rm{DD}} = \sideangle/2\pi$, respectively.

Considering that UEs employ omni-directional antennas, the sum of interference from all of the uplink transmissions, i.e., $I_{\mathrm{U}}$, can be expressed as
\vspace{-0mm}
\begin{equation}\label{eq:int_U}
I_{\mathrm{U}} = P_{\mathrm{U}} \sum_{u      \in \{{\Phi}_{\mathrm{U}}^{\mathrm{F,U}}       \cup       {\Phi}_{\mathrm{U}}^{\mathrm{H,U}} \}}\: K_{1}^{-\epsilon }Z_{u}^{\epsilon \alpha_1} {h_u K_{2}D_u^{-\alpha_{2}}},
\vspace{-0mm}
\end{equation}
where  $Z_{u}$ is the distance between the uplink UE $u$ and its serving \ac{BS}, $D_u$ is the distance between the uplink UE $u$ and the UE of interest, and $P_{\mathrm{U}} K_{1}^{-\epsilon }Z_{u}^{\epsilon \alpha_1}$ is the transmit power of the the uplink UE $u$. The symbol $h_u$ denotes the Rayleigh fading for the channel between the $u$th uplink UE and the UE of interest. 

By using (\ref{eq:sinr_fd_dwn}) and (\ref{eq:p_rx}), we can express the \ac{SINR} \ac{CCDF} for a given link distance $R$ as
\begin{equation*}
\begin{aligned}
& \mathrm{P}[\gamma_{\mathrm{FD,UE}} > y|r=R] =  \mathrm{P} \left[ \frac{P_{\mathrm{B}} \antboresightgain g_{b_{0}} K_{1}R^{-\alpha_{1}}}{N_{0}+I_{\mathrm{D}}+I_{\mathrm{U}}}  > y \right] \\
& = \mathrm{P} \left[ g_{b_{0}} > y P_{\mathrm{B}}^{-1}\antboresightgain^{-1} K_{1}^{-1} R^{\alpha_{1}} (N_{0}+I_{\mathrm{D}}+I_{\mathrm{U}}) \right] \\
\end{aligned}
\end{equation*}
\begin{equation}\label{eq:sinr_ccdf_2}
\overset{(a)} = e^{-\mu y P_{\mathrm{B}}^{-1} \antboresightgain^{-1} K_{1}^{-1} R^{\alpha_{1}} N_{0}}~\mathcal{L}_{I_{\mathrm{D}} + I_{\mathrm{U}}}(\mu y P_{\mathrm{B}}^{-1} \antboresightgain^{-1} K_{1}^{-1} R^{\alpha_{1}}),
\end{equation}
where (a) follows from the fact that $g_{b_{0}}$ $\sim$ $\exp(\mu )$. The Laplace transform of the total interference can be written as
\begin{equation}\label{eq:laplace_FD_D_1}
\begin{aligned}
\mathcal{L}_{I_{\mathrm{D}}+ I_{\mathrm{U}}}(s) = 
\mathbb{E}_{\Phi_{\mathrm{B}}^{\mathrm{F}}       \cup       \Phi_{\mathrm{B}}^{\mathrm{D}}       \cup      { \Phi}_{\mathrm{U}}^{\mathrm{F,U}} \cup      { \Phi}_{\mathrm{U}}^{\mathrm{H,U}}, g_b, h_u, Z_{u}, c_{\rm{dd}}}  \Bigg[ e^{-s (I_{\mathrm{D}}+I_{\mathrm{U}})} \Bigg],
\end{aligned}
\end{equation}
where  $s = \mu y P_{\mathrm{B}}^{-1} \antboresightgain^{-1} K_{1}^{-1} R^{\alpha_{1}}$ and $c_{\rm{dd}} \in \mathbb{C}^{\rm{DD}} = \{\rm{C}_1^{\rm{DD}}, \rm{C}_2^{\rm{DD}}   \}$. Since $\Phi_{\mathrm{B}}^{\mathrm{F}}$, $\Phi_{\mathrm{B}}^{\mathrm{D}}$, ${\Phi}_{\mathrm{U}}^{\mathrm{F,U}}$, and ${\Phi}_{\mathrm{U}}^{\mathrm{H,U}}$ are assumed to be independent \acp{PPP}, we rewrite \eqref{eq:laplace_FD_D_1} as
\begin{equation}\label{eq:laplace_FD_D_2}
\begin{aligned}
 &\mathcal{L}_{I_{\mathrm{D}} + I_{\mathrm{U}}}(s) = \\
 &\underbrace{\mathbb{E}_{\Phi_{\mathrm{B}}^{\mathrm{F}}       \cup       \Phi_{\mathrm{B}}^{\mathrm{D}}, g_b, c_{\rm{dd}}} \Bigg[ e^{-s I_{\mathrm{D}}}\Bigg]}_{L_x(s)} \times  \underbrace{\mathbb{E}_{{\Phi}_{\mathrm{U}}^{\mathrm{F,U}}      \cup       {\Phi}_{\mathrm{U}}^{\mathrm{H,U}}, h_u, Z_u}\Bigg[e^{-s I_{\mathrm{U}}} \Bigg]}_{L_y(s)}.
\end{aligned}
\end{equation}

By taking the expectation over $c_{\rm{dd}}$, the first term in \eqref{eq:laplace_FD_D_2} can be written as
\begin{equation}\label{eq:laplace_FD_D_3_A}
\begin{aligned}
 & L_x(s) = \\
 & \sum_{i=1}^2 p_i^{\rm{DD}} \mathbb{E}_{\Phi_{\mathrm{B}}^{\mathrm{F}}       \cup       \Phi_{\mathrm{B}}^{\mathrm{D}}, g_b} \Bigg[ e^{ -s \sum_{b      \in \{\Phi_{\mathrm{B}}^{\mathrm{D}}       \cup       \Phi_{\mathrm{B}}^{\mathrm{F}} \backslash{b_0}\}}  a_i^bP_{\rm B} g_b K_{1}R_b^{-\alpha_{1}} } \Bigg],
\end{aligned}
\end{equation}
where $a_i^b = G_b^{\rm{DD}}$ when $C_i^{\rm{DD}}$ holds.
By applying the Probability Generating Functional (PGFL) \cite{haenggi2013stochastic} of the \ac{PPP} to (\ref{eq:laplace_FD_D_3_A}), it can be further written as:
\begin{equation}\label{eq:laplace_FD_D_4_A}
\begin{aligned}
L_x(s) = \sum_{i=1}^2 p_i^{\rm{DD}} e^ { -2\pi \lambda_{\mathrm{B}}( \rho_{\mathrm{F}} + \rho_{\mathrm{D}} ) \int_{R}^{\infty} \left(\frac{s a_i^b K_{1} P_{\mathrm{B}} v^{-\alpha_{1}}}{s a_i^b K_{1} P_{\mathrm{B}} v^{-\alpha_{1}} + \mu } \right) v \mathrm{d}v }.
\end{aligned}
\end{equation}

By following the similar steps, the second term in (\ref{eq:laplace_FD_D_2}), i.e., $L_y(s)$, can be written as
\begin{equation}\label{eq:laplace_FD_D_5}
\begin{aligned}
&L_y(s) = \\
& e^{ -2\pi (\rho_{\mathrm{F}} + \rho_{\mathrm{U}}) \lambda_{\mathrm{B}} \int_{0}^{\infty}  \Bigg(1 - \mathbb{E}_{Z_u}\left[\frac{\mu }{s K_{2} P_{\mathrm{U}} K_1^{-\epsilon } Z_u^{\epsilon \alpha_1} v^{-\alpha_{2}} + \mu }\right] \Bigg) v \mathrm{d}v}.
\end{aligned}
\end{equation}

It is worth noting that the lower extreme of integration in (\ref{eq:laplace_FD_D_4_A}) is $R$ as the distance between the closest interfering \ac{BS} and the UE of interest is greater than $R$. However, the closest interfering uplink UE of an FD cell can also be in its own cell, so the lower extreme of integration in (\ref{eq:laplace_FD_D_5}) becomes $0$.
Assuming that there is no power control, i.e., {$\epsilon = 0$}, (\ref{eq:laplace_FD_D_5}) can be rewritten as
\begin{equation}\label{eq:laplace_FD_D_6_A}
L_y(s) = e^{-2\pi \lambda_{\mathrm{B}} (\rho_{\mathrm{F}} + \rho_{\mathrm{U}}) \int_{0}^{\infty} \left(\frac{s K_{2} P_{\mathrm{U}} v^{-\alpha_{2}}}{s K_{2} P_{\mathrm{U}} v^{-\alpha_{2}} + \mu } \right) v \mathrm{d}v }.
\end{equation}

Finally, we obtain the CCDF of the downlink SINR in a FD cell for a mixed system,
\begin{equation}\label{eq:sinr_downlink_finally}
\begin{split}
 &\mathrm{P}[\gamma_{FD,UE} > y] = \int_0^{\infty} \mathrm{P}[\gamma_{FD,UE} > y|r=R] f_r(R) \mathrm{d}R \\ =
 & \int_0^{\infty} e^{-s N_{0}}~\mathcal{L}_{x}(s)~\mathcal{L}_{y}(s) f_r(R) \mathrm{d}R, \nonumber
 \end{split}
\end{equation}
where $s = \mu y P_{\mathrm{B}}^{-1} \antboresightgain^{-1} K_{1}^{-1} R^{\alpha_{1}}$ and $f_r(R)$ is given by
\begin{equation}\label{eq:distance_dis_pdf_gen}
f_r(R) = e^{-\pi \lambda R^2} 2 \pi \lambda R,
\end{equation}
as the UE of interest is associated with the nearest \ac{BS} and the \ac{BS} deployment follows \ac{PPP} \cite{haenggi2013stochastic, Andrews_classic}. 
\subsection{Uplink SINR Distribution in an FD Cell}\label{sec:FDuplinkSINR}
The uplink SINR for the \ac{BS} of interest in a FD cell of the mixed system is given by
\vspace{-0mm}
\begin{equation}\label{eq:sinr_fd_up}
\gamma_{\mathrm{FD,BS}}=\frac{P_{\mathrm{RX,BS}}}{N_{1}+\widetilde{I}_{\mathrm{D}}+\widetilde{I}_{\mathrm{U}}+ I_{\rm{SI}}},
\vspace{-0mm}
\end{equation}
where $N_1$ is the noise power at the \ac{BS}, $\widetilde{I}_{\mathrm{D}}$ and $\widetilde{I}_{\mathrm{U}}$ are the total interference received at the \ac{BS} from all other downlink transmissions and from all the uplink transmissions, respectively, $I_{\rm{SI}}$ represents the residual self-interference due to being in \ac{FD} mode, which is discussed in Section~\ref{sec:selfInterference}, and $P_{\mathrm{RX,BS}}$ is the received signal power from the uplink UE. $P_{\mathrm{RX,BS}}$ can be expressed as
\begin{equation}\label{eq:p_rx_up}
P_{\mathrm{RX,BS}} = P_{\mathrm{U}} \antboresightgain h'_{u_{0}} K_{1}^{(1-\epsilon )}r^{\alpha_{1}(\epsilon-1)},
\end{equation}
where $r$ is the link distance between the \ac{BS} of interest and its uplink UE, and $h'_{u_{0}}$ denotes the Rayleigh fading for this link. $\widetilde{I}_{\mathrm{D}}$ and $\widetilde{I}_{\mathrm{U}}$ can be expressed as
\begin{equation}\label{eq:int_D_up}
\widetilde{I}_{\mathrm{D}} =  \sum_{b      \in \{\Phi_{\mathrm{B}}^{\mathrm{D}}       \cup       \Phi_{\mathrm{B}}^{\mathrm{F}} \backslash{b_0}\}}\: G_b^{\rm{DU}} P_{\mathrm{B}} g'_b K_{3}L_b^{-\alpha_{3}},
\end{equation}
and
\begin{equation}\label{eq:int_U_up}
\widetilde{I}_{\mathrm{U}} = \sum_{u      \in \{{\Phi}_{\mathrm{U}}^{\mathrm{F,U}}       \cup       {\Phi}_{\mathrm{U}}^{\mathrm{H,U}} \}: X_u > Z_u } G_u^{\rm{UU}} P_{\mathrm{U}} h'_u K_{1}^{(1-\epsilon )} Z_u^{\epsilon \alpha_1}X_u^{-\alpha_{1}}~,
\end{equation}
respectively, where $L_b$ and $X_u$ are the distance between the interfering \ac{BS}  $b$  and the \ac{BS} of interest and the distance between the interfering uplink UE $u$ and the \ac{BS} of interest, respectively; $Z_u$ is the distance between the interfering uplink UE $u$ and its serving \ac{BS}. It is worth noting that we introduce the condition $\{X_u > Z_u \}$ in~(\ref{eq:int_U_up}) for all $u   \in \{{\Phi}_{\mathrm{U}}^{\mathrm{F,U}}       \cup       {\Phi}_{\mathrm{U}}^{\mathrm{H,U}} \}$ as it guarantees that the distance $Z_u$ of the interfering UE $u$ to its serving \ac{BS} is shorter than the distance from $u$ to the victim \ac{BS}, which is also taken into account in  ~\cite{dynamic_TDD}. In (\ref{eq:int_D_up}), $G_b^{\rm{DU}}$ is the effective antenna gain given by 
\begin{equation}\label{eq:int_D_up_def}
\small{
	G_b^{\rm{DU}} = 
	\left \{ \,
		\begin{IEEEeqnarraybox}[] [c] {l?s}
			\IEEEstrut
			 \antboresightgain^2 &  if $\rm{C}^{\rm{DU}}_1$  holds, \\
             \antboresightgain \antsidegain &  if $\rm{C}^{\rm{DU}}_2$  holds, \\
             \antsidegain^2  &  if $\rm{C}^{\rm{DU}}_3$  holds, 
			\IEEEstrut
		\end{IEEEeqnarraybox}  
	\right. 
	}
\end{equation}
where $\rm{C}^{\rm{DU}}_1$, $\rm{C}^{\rm{DU}}_2$, and $\rm{C}^{\rm{DU}}_3$ are the condition that the receiving beam of the \ac{BS} of interest and the transmit beam of the interfering \ac{BS} $b$ are aligned to each other, the condition that either receiving beam of the \ac{BS} of interest or the transmit beam of the interfering \ac{BS} $b$ are aligned to each other, and the condition that neither receiving beam of the \ac{BS} of interest nor the transmit beam of the interfering \ac{BS} $b$ are aligned to each other, which occur with the probabilities of $p^{\rm{DU}}_1 = \boresightangle^2/4\pi^2$, $p^{\rm{DU}}_2 = 2\boresightangle\sideangle/4\pi^2$, and $p^{\rm{DU}}_3 = \sideangle^2/4\pi^2$, respectively. Similarly, in (\ref{eq:int_U_up}), the effective antenna gain in the uplink, i.e., $G_b^{\rm{UU}}$, can be defined as 
\begin{equation}\label{eq:int_U_up_def}
\small{
	G_u^{\rm{UU}} = 
	\left \{ \,
		\begin{IEEEeqnarraybox}[] [c] {l?s}
			\IEEEstrut
			 \antboresightgain  &  if $\rm{C}^{\rm{UU}}_1$  holds, \\
              \antsidegain  &  if $\rm{C}^{\rm{UU}}_2$  holds,  
			\IEEEstrut
		\end{IEEEeqnarraybox}
	\right. 
	}
\end{equation}
where $\rm{C}^{\rm{UU}}_1$ and $\rm{C}^{\rm{UU}}_2$  are the condition that the receiving beam of the \ac{BS} of interest is or is not oriented towards the interfering UE $u$, which occurs with the probability of $p^{\rm{UU}}_1 = \boresightangle/2\pi$ and with the probability of $p^{\rm{UU}}_2 = \sideangle/2\pi$, respectively. 

For a given link distance R, the \ac{SINR} \ac{CCDF} given in (\ref{eq:sinr_fd_up}) can be calculated as
\begin{equation}\label{eq:sinr_ccdf_u_2}
\begin{aligned}
&\mathrm{P}[\gamma_{\mathrm{FD,BS}} > y|r=R] = e^{-\mu y P_{\mathrm{U}}^{-1} \antboresightgain^{-1} K_{1}^{(\epsilon-1)} R^{\alpha_{1}(1-\epsilon )}N_{1}} \\
 &   \times    \mathcal{L}_{\widetilde{I}_{\mathrm{D}} + \widetilde{I}_{\mathrm{U}} + I_{\rm{SI}}}(\mu y P_{\mathrm{U}}^{-1} \antboresightgain^{-1} K_{1}^{(\epsilon-1)} R^{\alpha_{1}(1-\epsilon )}),
\end{aligned}
\end{equation}
where the Laplace transform of $(\widetilde{I}_{\mathrm{D}} + \widetilde{I}_{\mathrm{U}} + I_{\rm{SI}})$ is given by
\begin{equation}\label{eq:laplace_FD_U_1}
\begin{aligned}
\mathcal{L}_{\widetilde{I}_{\mathrm{D}} + \widetilde{I}_{\mathrm{U}} + I_{\rm{SI}}}(s) = \underbrace{\mathbb{E}_{\Phi_{\mathrm{B}}^{\mathrm{F}}       \cup       \Phi_{\mathrm{B}}^{\mathrm{D}}, g'_b, c_{\rm{du}}} \Bigg[ e^{-s \widetilde{I}_{\mathrm{D}}} \Bigg]}_{H_x(s)}    \times   \\
\underbrace{\mathbb{E}_{{\Phi}_{\mathrm{U}}^{\mathrm{F,U}}      \cup       {\Phi}_{\mathrm{U}}^{\mathrm{H,U}}, h'_u, Z_u,c_{\rm{uu}}}\Bigg[e^{-s \widetilde{I}_{\mathrm{U}}} \Bigg]}_{H_y(s)} \times \underbrace{\mathbb{E}_{\orientationTX, \orientationRX}\Bigg[e^{-s I_{\rm{SI}}} \Bigg]}_{H_z(s)},
\end{aligned}
\end{equation}
where $s=\mu y P_{\mathrm{U}}^{-1} \antboresightgain^{-1} K_{1}^{(\epsilon-1)} R^{\alpha_{1}(1-\epsilon)}$, $c_{\rm{du}} \in \mathbb{C}^{\rm{DU}} = \{\rm{C}_1^{\rm{DU}}, \rm{C}_2^{\rm{DU}}, \rm{C}_3^{\rm{DU}}  \}$ and $c_{\rm{uu}} \in \mathbb{C}^{\rm{UU}} = \{\rm{C}_1^{\rm{UU}}, \rm{C}_2^{\rm{UU}}, \rm{C}_3^{\rm{UU}}  \}$. 

The first term in (\ref{eq:laplace_FD_U_1}) can be written as
\begin{equation}\label{eq:laplace_FD_U_2A}
\begin{aligned}
&H_x(s) = \\
& \sum_{i=1}^3 p_i^{\rm{DU}} \mathbb{E}_{\Phi_{\mathrm{B}}^{\mathrm{F}}       \cup       \Phi_{\mathrm{B}}^{\mathrm{D}}, g'_b} \Bigg[ e^{-s  \sum_{b      \in \{\Phi_{\mathrm{B}}^{\mathrm{D}}       \cup       \Phi_{\mathrm{B}}^{\mathrm{F}} \backslash{b_0}\}}\: {f^b_i g'_b P_{\mathrm{B}} K_{3}L_b^{-\alpha_{3}}}} \Bigg] \\
&= \sum_{i=1}^3 p_i^{\rm{DU}} e^{-2\pi (\rho_{\mathrm{F}} + \rho_{\mathrm{D}}) \lambda_{\mathrm{B}} \int_{0}^{\infty} \left(\frac{s f^b_i K_{3} P_{\mathrm{B}} v^{-\alpha_{3}}}{s f^b_i K_{3} P_{\mathrm{B}} v^{-\alpha_{3}} + \mu } \right) v \mathrm{d}v}~,
\end{aligned}
\end{equation}
where $f_i^b = G_b^{\rm{DU}}$ when $C_i^{\rm{DU}}$ holds true.  The lower extreme of integration in~(\ref{eq:laplace_FD_U_2A}) is zero as the closest interferer \ac{BS} (either FD or \ac{HD}) can be at any distance greater than $0$. 

The second term in (\ref{eq:laplace_FD_U_1}) can be written as
\begin{equation}\label{eq:laplace_FD_U_3}
\begin{aligned}
 &H_y(s) = \sum_{i=1}^2 p_i^{\rm{UU}} \times\\
& \mathbb{E}_{{\Phi}_{\mathrm{U}}^{\mathrm{F,U}}      \cup       {\Phi}_{\mathrm{U}}^{\mathrm{H,U}}, h'_u, Z_u}\Bigg[e^{-s \sum_{u      \in \{{\Phi}_{\mathrm{U}}^{\mathrm{F,U}}       \cup       {\Phi}_{\mathrm{U}}^{\mathrm{H,U}}\}: X_u > Z_u }\: {\frac{ t^b_i h'_u P_{\mathrm{U}} Z_u^{\epsilon \alpha_1}}{K_{1}^{(\epsilon-1)} X_u^{\alpha_1}}}} \Bigg] \\
& = \sum_{i=1}^2 p_i^{\rm{UU}} \times \\
&e^{-2\pi (\rho_{\mathrm{F}} + \rho_{\mathrm{U}}) \lambda_{\mathrm{B}} \int_{0}^{\infty}  \left(1 - \mathbb{E}_{Z_u}\left[\frac{\mu }{ \frac{s t^b_i P_{\mathrm{U}} Z_u^{\epsilon \alpha_1} \mathds{1}\{Z_u < v\}}{K_1^{\epsilon-1 }  v^{\alpha_{1}}}  + \mu }\right] \right) v \mathrm{d}v }~,
\end{aligned}
\end{equation}
where $t_i^b = G_u^{\rm{UU}}$ when $C_i^{\rm{UU}}$ holds true.
The lower extreme of integration in~(\ref{eq:laplace_FD_U_3}) is also zero but the constraint $\{Z_u < v\}$ makes sure that only UEs from the other cells are included in the interference term. The integration in (\ref{eq:laplace_FD_U_3}) can be further simplified using integration by parts as given in~\cite{SGoyal_ICC_Arxiv}. When there is no power control ($\epsilon = 0$), it can be written as:
\begin{equation}\label{eq:laplace_FD_U_4A}
\begin{aligned}
&H_y(s) = \\
&  \sum_{i=1}^2 p_i^{\rm{UU}} e^{ -2\pi  ( \rho_{\mathrm{F}} + \rho_{\mathrm{U}} ) \lambda_{\mathrm{B}} \int_{0}^{\infty}  \left( \frac{s  t^b_i P_{\mathrm{U}} K_1 v^{-\alpha_1}}{\mu + s  t^b_i P_{\mathrm{U}} K_1 v^{-\alpha_1}}\right) \mathbb{P} (Z_u \leq v) v \mathrm{d}v}, \nonumber
\end{aligned}
\end{equation}
where $\mathbb{P} (Z_u \leq v) $ is assumed to be $\mathbb{P}\{Z_u \leq v\} = 1-\mathrm{exp}(-\pi \nu \lambda_{\mathrm{B}} v^2)$, and $\nu\geq 0$ is a correction factor that takes into account the effect of the correlation among points on the distance distribution. It has been shown that $\nu = 1.25$ provides well-aligned results \cite{SGoyal_ICC_Arxiv}. 

The third term in (\ref{eq:laplace_FD_U_1}) can be written as
\begin{equation}\label{eq:laplace_self_int}
\begin{aligned}
H_z(s) = \sum_{i=1}^4 p_i^{\rm{SI}} e^{-sw_i},
\end{aligned}
\end{equation}
where $w_i = I_{\rm{SI}}$ such that condition $C_i^{\rm{SI}}$ holds. The probability $p_i^{\rm{SI}}$ can be derived based on the conditions defined in Section~\ref{sec:selfInterference}. 

Finally, we obtain the CCDF of the uplink SINR in a FD cell for a mixed system as
\begin{equation}\label{eq:sinr_ccdf_u_finally}
\begin{split}
&\mathrm{P}[\gamma_{\mathrm{FD,BS}} > y] =
\int_0^{\infty} \mathrm{P}[\gamma_{\mathrm{FD,BS}} > y|r=R] f'_r(R) \mathrm{d}R=
\\
 &\int_0^{\infty} e^{-s N_{1}} \mathcal{L}_{\widetilde{I}_{\mathrm{D}} + \widetilde{I}_{\mathrm{U}} + I_{\rm{SI}}}(s) f'_r(R) dR, \nonumber
\end{split}
\end{equation}
where $s=\mu y P_{\mathrm{U}}^{-1} \antboresightgain^{-1} K_{1}^{(\epsilon-1)} R^{\alpha_{1}(1-\epsilon)}$ and $f'_r(R)$ is given by
\begin{equation}\label{eq:distance_dis_pdf_mod}
f'_r(R) = e^{-\pi  \nu \lambda_{\mathrm{B}} R^2} 2 \pi \nu \lambda_{\mathrm{B}} R~.
\end{equation}

\subsection{Downlink and Uplink SINR Distribution in an HD Cell}\label{sec:HDdownlinkSINR}
The downlink SINR at a UE in a \ac{HD} cell of the mixed system can be derived similarly to the downlink SINR in a FD cell. A downlink UE in a \ac{HD} cell gets interference from all simultaneous uplink and downlink transmissions similar to the downlink UE in a FD cell. However, there is one difference from the derivation of the SINR CCDF given in Section~\ref{sec:FDdownlinkSINR}. To consider the interference from all the active uplink transmissions, the lower extreme of integration in (\ref{eq:laplace_FD_D_5}) is zero, which includes the uplink transmission in its own FD cell, whereas in the case of a \ac{HD} cell, we need to make sure that no uplink transmission inside the downlink UE's own cell is included. For analytical tractability, to take this into account, we make an approximation that the distance from the nearest interfering uplink transmission is approximated by the distance from the nearest interfering \ac{BS}. This is the same approximation made in \cite{Quekhybrid, elsawy_alpha_duplex} while modeling the UE-to-UE interference at a FD UE. Thus, in this case, the lower extreme of integration in (\ref{eq:laplace_FD_D_5}) will be $R$, i.e., the distance of the downlink UE from its serving \ac{BS}. For this case,
\vspace{-0mm}
\begin{equation}\label{eq:laplace_FD_D_5_downlink_HD}
\begin{aligned}
&L'_y(s) =  \\
&e^{ -2\pi (\rho_{\mathrm{F}} + \rho_{\mathrm{U}}) \lambda_{\mathrm{B}} \int_{R}^{\infty}  \Bigg(1 - \mathbb{E}_{Z_u}\left[\frac{\mu }{s K_{2} P_{\mathrm{U}} K_1^{-\epsilon } Z_u^{\epsilon \alpha_1} v^{-\alpha_{2}} + \mu }\right] \Bigg) v \mathrm{d}v} .
\end{aligned}
\end{equation}
Similar to~(\ref{eq:sinr_downlink_finally}), the expression for CCDF of SINR in \ac{HD} cell is given by,
\vspace{-0mm}
\begin{equation}\label{eq:sinr_hd_down_CCDF}
\begin{aligned}
\mathrm{P}[\gamma_{\mathrm{HD,UE}} > y] = \int_0^{\infty} e^{-s N_{0}}~L_x(s)~L'_y(s) f_r(R) \mathrm{d}R,
\end{aligned}
\end{equation}
where $s = \mu y P_{\mathrm{B}}^{-1} \antboresightgain^{-1} K_{1}^{-1} R^{\alpha_{1}}$.

In the uplink case, the expression for uplink SINR in a \ac{HD} cell will be the same as the uplink SINR in a FD cell in Section~\ref{sec:FDuplinkSINR} but without any self-interference, i.e.,~$I_{\rm{SI}} = 0$,
\vspace{-0mm}
\begin{equation}\label{eq:sinr_hd_up}
\gamma_{\mathrm{HD,BS}}=\frac{P_{\mathrm{RX,BS}}}{N_{1}+\widetilde{I}_{\mathrm{D}}+\widetilde{I}_{\mathrm{U}}}.
\end{equation}
The CCDF of $\gamma_{\mathrm{HD,BS}}$ is given by,
\begin{equation}\label{eq:sinr_ccdf_u_1_hd}
\mathrm{P}[\gamma_{\mathrm{HD,BS}} > y] = \int_0^{\infty} \mathrm{P}[\gamma_{\mathrm{HD,BS}} > y|r=R] f'_r(R) \mathrm{d}R,
\end{equation}
where
\begin{equation}\label{eq:sinr_ccdf_u_2_hd}
\begin{aligned}
\mathrm{P}[\gamma_{\mathrm{HD,BS}} > y|r=R] = e^{-s N_{1}} H_x(s) H_y(s),
\end{aligned}
\end{equation}
with $ s = \mu y P_{\mathrm{U}}^{-1} \antboresightgain^{-1} K_{1}^{(\epsilon-1)} R^{\alpha_{1}(1-\epsilon )}$, where, $H_x(s)$, and $H_y(s)$ are given in (\ref{eq:laplace_FD_U_2A}), and (\ref{eq:laplace_FD_U_3}), respectively.

\section{Average rate}\label{sec:avg_rate}
The average rate per hertz can be computed as \cite{jeffAndrews_uplink,Goyal_arxiv}
\begin{align}
\mathbb{E}[\mathrm{C}] = \int
_{0}^{\infty}\mathrm{P}\left[\log_{2}(1+\gamma )>u\right]\mathrm{d}u.
\end{align}

By using (\ref{eq:sinr_ccdf_2}), the average downlink rate in an FD cell is given by
\begin{equation}\label{eq:rate_donwlink_FD}
\begin{aligned}
&\mathbb{E}[\mathrm{C_{\rm FD,UE}}] = \int_{0}^{\infty}\mathrm{P}\left[\log_{2}(1+\gamma_{\rm FD,UE})>u\right]\mathrm{d}u \\
&= \int _{0}^{\infty}\int_{0}^{\infty}  e^{-\mu (2^{u}-1) P_{\mathrm{B}}^{-1} \antboresightgain^{-1} K_{1}^{-1} R^{\alpha_{1}} N_{0}}    \times   \\
&\mathcal{L}_{I_{\mathrm{D}} + I_{\mathrm{U}}}(\mu (2^{u}-1) P_{\mathrm{B}}^{-1} \antboresightgain^{-1} K_{1}^{-1} R^{\alpha_{1}})f_{r}(R)~\mathrm{d}R~\mathrm{d}u.
\end{aligned}
\end{equation}

By using (\ref{eq:sinr_ccdf_u_2}), the average uplink rate in a FD cell is given by
\begin{equation}\label{eq:rate_uplink_FD}
\begin{aligned}
&\mathbb{E}[\mathrm{C_{FD,BS}}] = \int_{0}^{\infty}\mathrm{P}\left[\log_{2}(1+\gamma_{\rm FD,BS})>u\right]\mathrm{d}u \\
& = \int _{0}^{\infty}\int_{0}^{\infty}  e^{-\mu (2^{u}-1) P_{\mathrm{U}}^{-1} \antboresightgain^{-1} K_{1}^{(\epsilon-1)} R^{\alpha_{1}(1-\epsilon )} N_{1}}    \times    \\
&\mathcal{L}_{\widetilde{I}_{\mathrm{D}} + \widetilde{I}_{\mathrm{U}} + I_{\rm{SI}}}(\mu (2^{u}-1) P_{\mathrm{U}}^{-1} \antboresightgain^{-1} K_{1}^{-1} R^{\alpha_{1}})f'_{r}(R)~\mathrm{d}R~\mathrm{d}u. \nonumber
\end{aligned}
\end{equation}

Similarly the average downlink and uplink rates in a \ac{HD} cell, i.e, $\mathrm{E}[\mathrm{C_{HD,UE}}]$, $\mathrm{E}[\mathrm{C_{HD,BS}}]$, respectively, can be derived. Combining the rates of FD and \ac{HD} cells, the average downlink and uplink rates of the complete network are given by
\begin{equation}\label{eq:rate_donwlink_complete}
\mathbb{E}[\mathrm{C_{D}}] = \rho_F \mathbb{E}[\mathrm{C_{FD,UE}}] + \rho_D \mathbb{E}[\mathrm{C_{HD,UE}}]~,
\end{equation}
and
\begin{equation}\label{eq:rate_uplink_complete}
\mathbb{E}[\mathrm{C_{U}}] = \rho_F \mathbb{E}[\mathrm{C_{FD,BS}}] + \rho_U \mathbb{E}[\mathrm{C_{HD,BS}}]~,
\end{equation}
respectively. Hence, the downlink and uplink ASEs of the mixed network can be obtained from~(\ref{eq:rate_donwlink_complete}) and~(\ref{eq:rate_uplink_complete}), respectively, as $\mathrm{ASE_D} = \lambda_{\mathrm{B}}\mathbb{E}[\mathrm{C_{D}}]$ and $\mathrm{ASE_U} = \lambda_{\mathrm{B}}\mathbb{E}[\mathrm{C_{U}}]$.

\section{Numerical Results}\label{sec:num_res}

We evaluate the \ac{ASE} and coverage of a mixed \ac{HD} and \ac{FD} network in the case of directional transmission at the \acp{BS}. We also provide results with traditional TDD \ac{HD} (THD) systems, in which, all the \acp{BS} are involved only in \ac{HD} operations, i.e., a \acp{BS} schedules either uplink or downlink transmission. We simulate different THD systems while varying the proportion of cells in downlink and uplink transmissions. The network parameters are tabulated in Table~\ref{tab:simulation_parameters}. In the case of directional transmissions at the \acp{BS}, we consider two different antenna patterns listed as: 1) $\theta_{\rm{B}}$= 35$^\circ$ with $G_{\rm{B}}$ = 15 dBi, $G_{\rm{S}}$ = 0 dBi, 2) $\theta_{\rm{B}}$= 90$^\circ$ with $G_{\rm{B}}$ = 7 dBi, $G_{\rm{S}}$ = 0 dBi \cite{3GPP:5}.
\begin {table}
\caption {Network Parameters} \label{tab:simulation_parameters}
\vspace{0mm}
\begin{center}
    \begin{tabular}{| p{1.5 in} | p{1.2 in} |}
    	\hline
		\textbf {Parameter} & \textbf{Value} \\ \hline
		Bandwidth & $10$ MHz \\ \hline
		\ac{BS} Density [nodes/m$^2$]  & $10^{-3}$  \\ \hline
		Thermal Noise Density & $-174$ dBm/Hz \\ \hline
        Outage SINR Threshold & $-8$ dB \\ \hline
		Noise Figure & $9$ dB (UE), $8$ dB (\ac{BS})\\ \hline
        $K_1$, $K_2$, $K_3$ \cite{3GPP36814} & 8.8 $\times$ $10^{-4}$\\ \hline
        $\alpha_1$, $\alpha_2$, $\alpha_2$ \cite{3GPP36814} & 3.67 \\ \hline
        $A$ & 120 dB \\ \hline
        $\epsilon$  & 0 \\ \hline
    \end{tabular}
\end{center}
\vspace{-6mm}
\end{table}

\begin{figure}[t]
\centering
\includegraphics[width = 3.4 in] {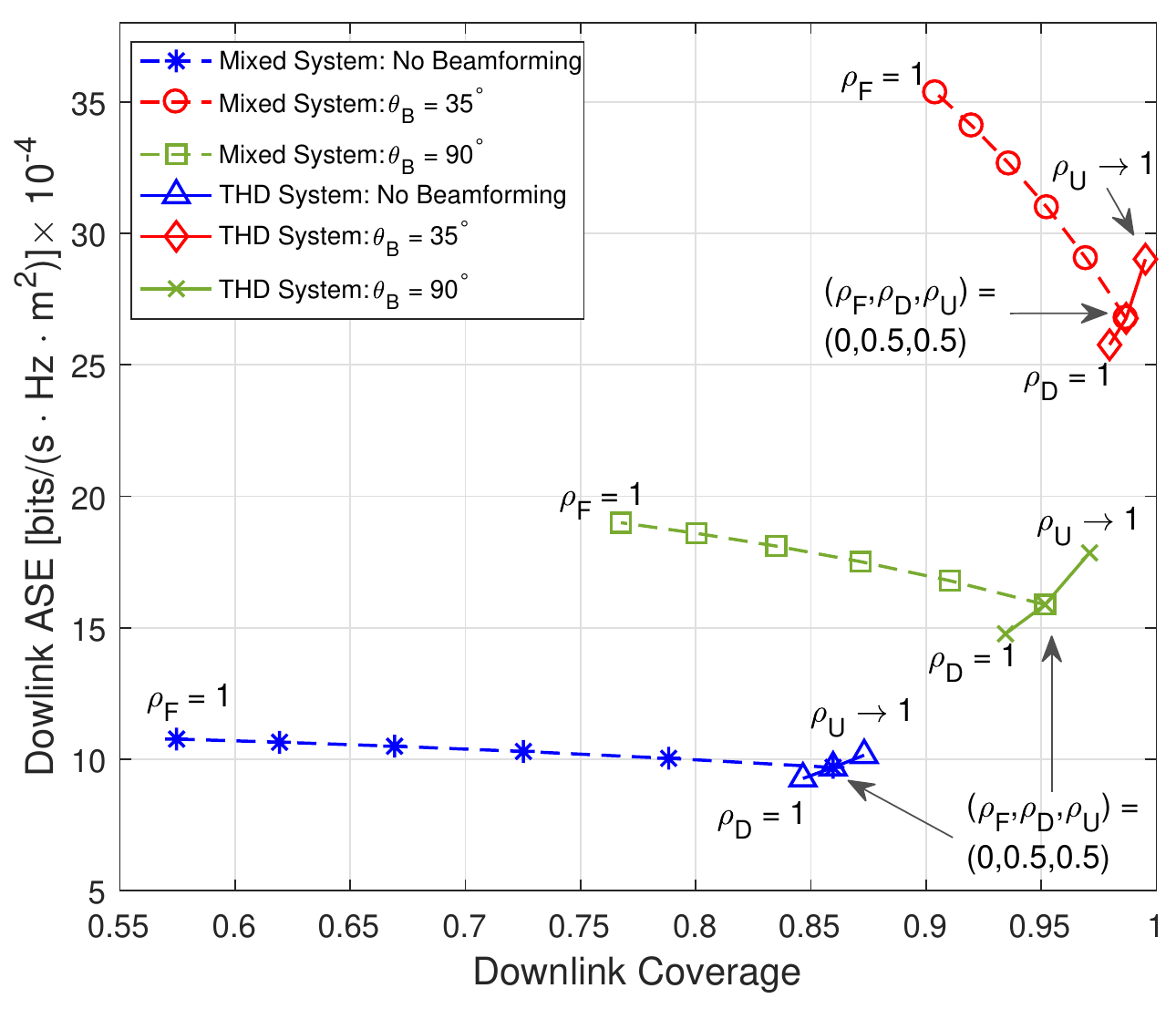}
\caption{Trade-off between ASE and coverage in downlink with directional transmissions.
}
\label{fig:down_coverage_ase_tradeoff}
\vspace{-5mm}
\end{figure}

\begin{figure}[t]
\centering
\includegraphics[width = 3.5 in] {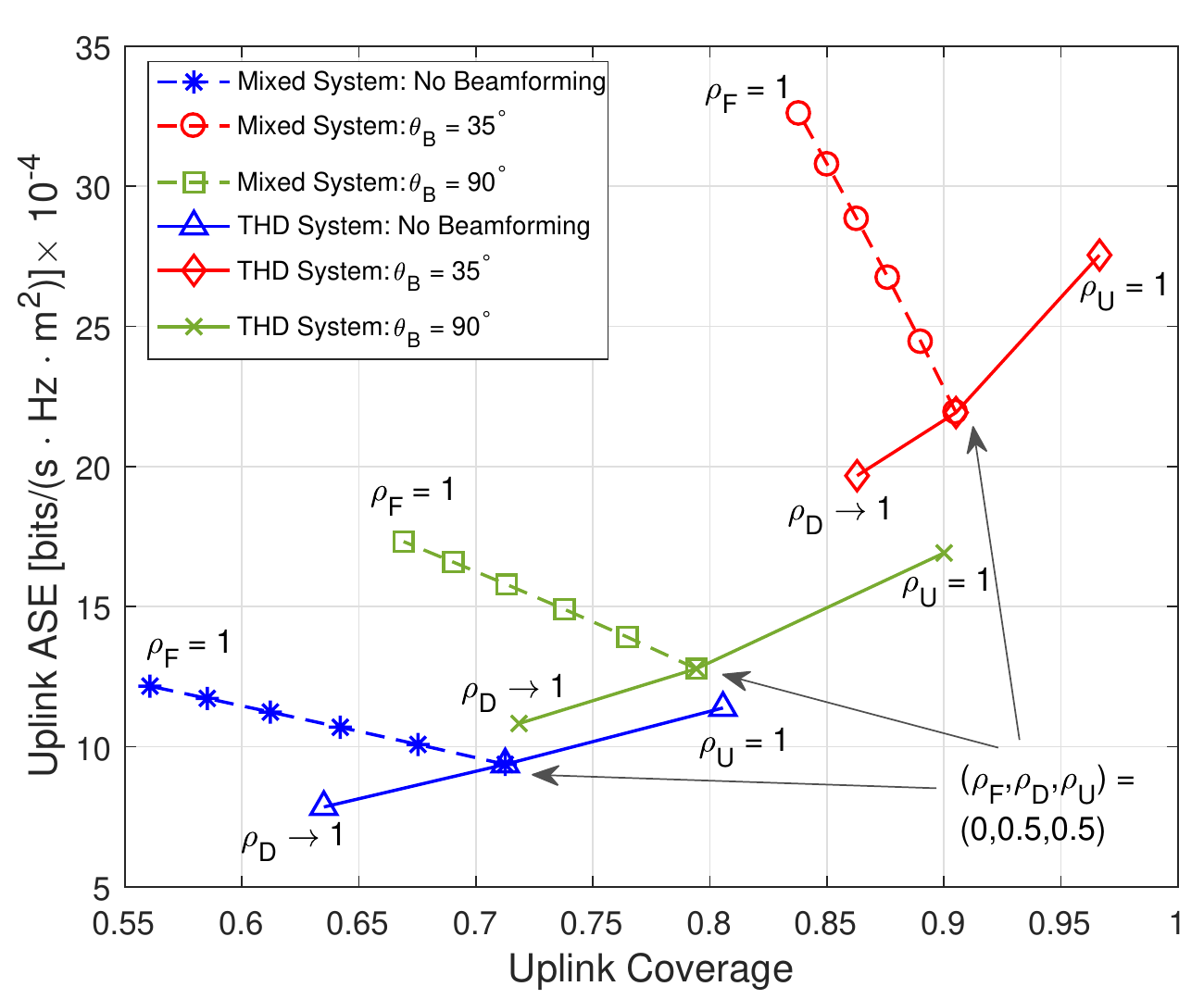}
\caption{Trade-off between ASE and coverage in uplink with directional transmissions. 
\vspace{-5mm}
}
\label{fig:up_coverage_ase_tradeoff}
\end{figure}

Figs.~\ref{fig:down_coverage_ase_tradeoff} and~\ref{fig:up_coverage_ase_tradeoff} show the ASE  vs. coverage trade-off in downlink and uplink for different antenna settings, respectively, when $P_{\mathrm{B}} = 24$ dBm and $P_{\mathrm{U}} = 23 $ dBm. It includes the performance of different mixed and THD systems. In the case of THD systems, $\rho_D = 1$, and $\rho_U = 1$ represent the scenarios of all the \acp{BS} scheduled in downlink and uplink transmissions, respectively. 
The overall coverage rates of the mixed system in downlink and uplink are computed as $(\rho_{\mathrm{F}}\Theta_{\mathrm{FD,DL}}+\rho_{\mathrm{D}} \Theta_{\mathrm{HD,DL}})/(\rho_{\mathrm{F}}+\rho_{\mathrm{D}})$ and  $(\rho_{\mathrm{F}}\Theta_{\mathrm{FD,UL}}+\rho_{\mathrm{D}} \Theta_{\mathrm{HD,UL}})/(\rho_{\mathrm{F}}+\rho_{\mathrm{D}})$, respectively, where $\Theta_{\mathrm{FD,DL}}$, $\Theta_{\mathrm{FD,UL}}$, $\Theta_{\mathrm{HD,DL}}$, and $\Theta_{\mathrm{HD,UL}}$ are the coverage of FD downlink, FD uplink, HD downlink, and \ac{HD} uplink in the mixed system, respectively. 
The coverage is defined as the fraction of UEs in a non-outage region, where an outage happens if the received SINR is below the outage SINR threshold. The trade-off in the mixed system is presented for a given percentage of FD \acp{BS}, i.e., $\rho_{\mathrm{F}}$. The remaining \acp{BS} in the mixed system are equally divided into \ac{HD} downlink and \ac{HD} uplink modes, i.e., $\rho_{\mathrm{D}} = \rho_{\mathrm{U}} = (1 - \rho_{\mathrm{F}})/2$. In all antenna configurations, by increasing the number of \acp{BS} in FD mode, both downlink and uplink ASE increase at the cost of lower coverage. This trade-off occurs as  the number of transmissions and the aggregated interference in each direction increase for higher $\rho_{\rm{F}}$.

For both mixed and the THD systems, beamforming at the \acp{BS} provides gain both in terms of ASE and coverage. For example, in the mixed system with $\rho_{\mathrm{F}}$ = 0.4, with increasing the beamforming gain, i.e., changing $\theta_{\rm{B}}$ from 90$^\circ$ to 35$^\circ$, the ASE increases by 77\% and 79\% in the downlink and the uplink while the downlink and uplink coverage rates are increased by 9\% and 19\%, respectively. Applying beamforming at the \acp{BS} provides beamforming gain to both the downlink and the uplink transmissions while reducing the interference from the other nodes. In the downlink, the interference from the neighboring \ac{BS}, and in the uplink the interference from both the UE and \ac{BS} of the neighboring cell, and the self-interference decrease. Beamforming provides gain to all the mixed and the THD systems. Moreover, with higher beamforming gain, as we increase the number of FD cells, the rate of increment in ASE is much higher than the rate of decrement in the coverage.

In the case of THD systems, increasing the number of transmissions in the downlink direction ($\rho_D \to 1$) reduces both ASE and coverage in both downlink and uplink directions. It is because a downlink transmission generates interference from a \ac{BS} which is generally stronger than the interference from a UE transmitting in the uplink direction. This also holds for the mixed system, where for lower values of $\rho_{\mathrm{F}}$, most of the cells are in \ac{HD} mode, including both uplink and downlink transmissions. Therefore, both downlink and uplink performances in the mixed system are superior to the THD systems consisting of higher number of downlink transmissions.

\section{Conclusion}\label{sec:conc}
In this paper, we investigate a mixed duplexing cellular system composed of FD and \ac{HD} cells with directional transmission at the \acp{BS}. We consider a stochastic geometry-based model to derive the SINR complementary CDF and the ASE for the downlink and uplink directions. We study the impact of FD cells on the ASE vs. coverage trade-off of mixed systems for different beamforming configurations at the \acp{BS}. We show that the beamforming at the \acp{BS} increases the performance of both mixed as well as the traditional \ac{HD} systems significantly in both uplink and downlink directions. With higher beamforming gain, as we increase the number of FD cells, the gain in ASE increases rapidly with a small loss in the coverage of the network. 
Further extensions to our study could include opportunistic scheduling, more comprehensive antenna patterns including residual interference models for FD communications, beamforming at the UE side, and operation in millimeter wave frequencies.

\bibliographystyle{IEEEtran}
\bibliography{FD_references}

\end{document}